\documentclass[aps,prl,twocolumn,floatfix,citeautoscript,longbibliography]{revtex4-1}

\usepackage{pdfpages}

\makeatletter
\AtBeginDocument{\let\LS@rot\@undefined}
\makeatother

\usepackage{amsmath,amssymb,amsfonts}
\usepackage{graphicx}
\usepackage{bm}

\newcommand{\ket}[1]{\lvert #1\rangle}
\newcommand{\bra}[1]{\langle#1 \rvert}
\newcommand{\abs}[1]{\lvert #1 \rvert}

\newcommand{\br}{\mathbf{r}}
\newcommand{\bk}{\mathbf{k}}

\newcommand{\bK}{\mathbf{K}}

\begin{document}

\title{Signatures of adatom effects in the quasiparticle spectrum of Li-doped graphene}

\author{Kristen Kaasbjerg}
\email{kkaasbjerg@gmail.com}
\author{Antti-Pekka Jauho}
\affiliation{Center for Nanostructured Graphene (CNG), Department of Physics, Technical University of Denmark, DK-2800 Kgs. Lyngby, Denmark}

\begin{abstract} We study the spectral function and quasiparticle scattering in
  Li-decorated graphene (Li@graphene) with an atomistic $T$-matrix formalism and
  uncover adatom-induced spectral effects which shed light on experimentally
  observed angle-resolved photoemission spectroscopy (ARPES) features. From
  transport studies, alkali adatoms are known to introduce charged-impurity
  scattering limiting the carrier mobility. Here, we demonstrate that Li adatoms
  furthermore give rise to a low-energy impurity band centered at the $\Gamma$
  point which originates from the hybridization between the atomic 2s state of
  the Li adatoms and graphene "surface" states. We show that the impurity band
  is strongly dependent on the concentration $c_\mathrm{Li}$ of Li adatoms, and
  aligns with the Li-induced Fermi level on the Dirac cone at
  $c_\mathrm{Li}\sim 8\,$\% ($E_F\approx 1.1$~eV). Finally, we show that
  adatom-induced quasiparticle scattering increases dramatically at energies
  above $\sim 1\,\mathrm{eV}$ close to the van Hove singularity in the graphene
  density of states (DOS), giving rise to a large linewidth broadening on the
  Dirac cone with a concomitant downshift and a characteristic kink in the
  conduction band. Our findings are highly relevant for future studies of ARPES,
  transport, and superconductivity in adatom-doped graphene.
\end{abstract}

\date{\today}
\maketitle

\textbf{\textit{Introduction.}}---Graphene decorated with metallic adatoms has
emerged as an interesting platform for engineering graphene's electronic
properties and realizing novel electronic phases such as, e.g., the quantum spin
Hall phase~\cite{Wu:Engineering,Roche:Multiple,Ferreira:Impact} and
superconductivity~\cite{Neto:Superconducting,Mauri:Phonon,Giustino:Two}. In
addition, doping with adatoms opens the opportunity to probe the electronic
properties of graphene at high energies with, e.g., Fermi levels in excess of
1~eV in alkali-doped
graphene~\cite{Hofmann:Elph,Rotenberg:Extended,Lanzara:Many,Gruneis:Observation,Damascelli:Evidence,Forti:Alkali}. However,
metallic adatoms at the same time introduce charged-impurity scattering, thus
limiting the transport in adatom-doped
graphene~\cite{Ishigami:Charge,Fuhrer:Correlated,Forti:Alkali,Eisenstein:Transport,Folk:Weak}.

The spectral properties of adatom-doped graphene have been studied with
angle-resolved photoemission spectroscopy (ARPES)~\cite{Shen:RMP} in several
works~\cite{Rotenberg:Quasiparticle,Hofmann:Elph,Rotenberg:Extended,Lanzara:Many,Lanzara:Charge,Gruneis:Observation,Ast:Long,Damascelli:Evidence},
demonstrating many-body effects such as, e.g., electron-electron and
electron-phonon (el-ph)
interactions~\cite{Mauri:ElPh,Louie:Velocity,Sarma:Phonon,Sarma:Quasi,MacDonald:Plasmons,Louie:Angle},
while the effect of adatom-induced impurity
scattering~\cite{Ishigami:Charge,Fuhrer:Correlated,Forti:Alkali,Eisenstein:Transport}
in ARPES is not well understood~\cite{Ast:Long}. In addition, signatures of
superconductivity have been observed in ARPES on Li-decorated graphene
(Li@graphene) at Li concentrations corresponding to a Fermi level of
$E_F\sim 1$~eV~\cite{Damascelli:Evidence}. They appeared alongside a spectral
feature at the $\Gamma$ point which in the ordered LiC$_6$
structure~\cite{Mauri:Phonon} and alkali metal-graphite intercalation
compounds~\cite{Simons:The,Mauri:Theoretical,Nair:Superconductivity} corresponds
to a metal-atom dominated band predicted to promote
superconductivity~\cite{Mauri:Phonon}.

\begin{figure}[!b] \centering
  \includegraphics[width=0.99\linewidth]{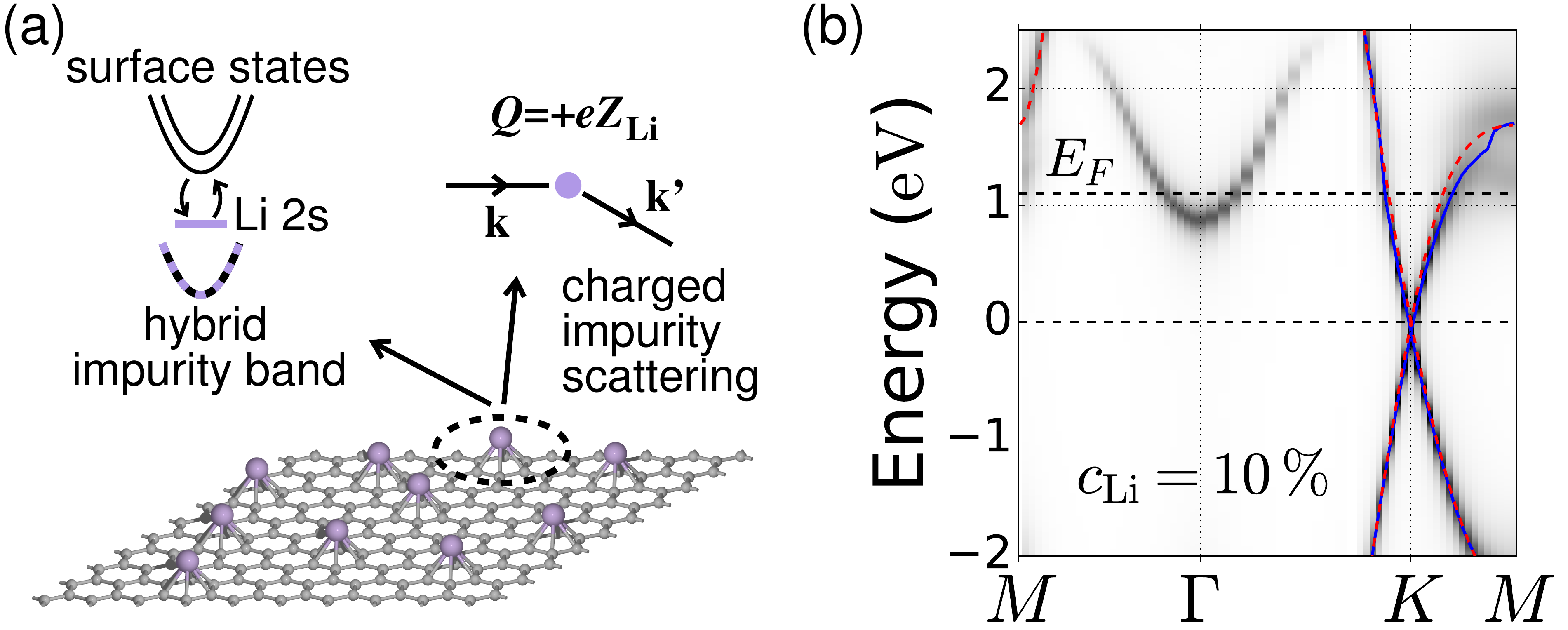}
  \caption{(a) Atomic illustration of disordered Li@graphene with a random
    configuration of Li adatoms giving rise to (i) a hybrid impurity band, and
    (ii) pronounced charged-impurity scattering as sketched in the top part. (b)
    Spectral function of Li@graphene at a Li concentration of
    $c_\mathrm{Li}=10$~\% showing the appearance of the $\Gamma$-point centered
    hybrid impurity band $\sim 0.9$~eV above the Dirac point ($E=0$). The dashed
    horizontal line indicates the Li-induced Fermi level $E_F$. The unperturbed
    Dirac-cone bands (red dashed lines) are renormalized (solid blue lines) and
    broadened by Li-induced impurity scattering.}
\label{fig:intro}
\end{figure}
In this work, we study the spectral properties of \emph{disordered} Li@graphene,
i.e., graphene with a random configuration of Li adatoms as illustrated in
Fig.~\ref{fig:intro}(a), using an atomistic first-principles $T$-matrix
formalism based on a parameter-free description of the impurity
potential~\cite{Jauho:Symmetry,Kaasbjerg:First}. This allows for a detailed
description of (i) the spectral $\Gamma$-point feature including its hitherto
unexplored concentration dependence as well as (ii) the \emph{renormalization}
and \emph{linewidth broadening} of the bands due to adatom-induced quasiparticle
(QP) scattering.

We find that in disordered Li@graphene, the $\Gamma$-point feature corresponds
to the hybrid impurity band shown in Fig.~\ref{fig:intro}(b), which originates
from the coupling between free-electron like "surface" states localized in
proximity to the graphene layer and the atomic Li 2s state as sketched in
Fig.~\ref{fig:intro}(a). The impurity band evolves downwards in energy from the
position of the Li 2s state with increasing $c_\mathrm{Li}$ and aligns with the
Fermi level on the Dirac cone at $c_\mathrm{Li}\sim 8\,\%$
($E_F\approx 1.1$~eV). In agreement with experiments~\cite{Damascelli:Evidence},
this suggests that conditions favorable for superconductivity may be realized in
disordered Li@graphene at concentrations well below the concentration
($c_\mathrm{Li}=33\,\%$) in LiC$_6$~\cite{Mauri:Phonon}.

We furthermore analyze the effect of adatom scattering on the Dirac cone states,
and find that charged-impurity scattering dominates the linewidth broadening,
while resonant
scattering~\cite{Basko:Resonant,Falko:Adsorbate,Katsnelson:Resonant,Fabian:Resonant}
by the atomic Li 2s state is negligible. Near the $M$ point, the linewidth
broadening in Fig.~\ref{fig:intro}(b) increases dramatically due to strong
charged-impurity scattering with a concomitant downshift and a kink in the QP
band [Fig.~\ref{fig:intro}(b) blue line]. Similar features have been observed at
high Fermi energies in different types of metal atom-doped
graphene~\cite{Rotenberg:Extended,Hofmann:Elph,Gruneis:Observation,Starke:Introducing},
and may yield an artificially high and anisotropic el-ph coupling if attributed
entirely to el-ph scattering as pointed out in
Refs.~\onlinecite{Louie:Van,Mauri:Electronic}. As we demonstrate, adatom-induced
impurity scattering presents a nonnegligible "intrinsic" contribution to the
band renormalization in adatom-doped graphene.

\textbf{\textit{Theory and methods.}}---For a random distribution of Li adatoms
as illustrated in Fig.~\ref{fig:intro}(a), the spectral properties probed in
ARPES are given by the impurity-averaged Green's function (GF). Here, we apply
the atomistic density-functional theory (DFT)-based $T$-matrix formalism described in
Refs.~\onlinecite{Jauho:Symmetry,Kaasbjerg:First} to calculate the GF for
disorderd Li@graphene.

The impurity-averaged Green's function is given by the Dyson
equation~\cite{Mahan,Flensberg}
\begin{equation}
    \label{eq:disavg} \hat{G}^{-1}_{\bk}(\varepsilon) =
\hat{G}_{\bk}^{0\;-1}(\varepsilon) - \hat{\Sigma}_{\bk}(\varepsilon) ,
\end{equation} where $\bk \in 1$st Brillouin zone (BZ) and the carets indicate a
matrix structure in the band index $n$. Impurity effects enter via the
self-energy $\hat{\Sigma}_{\bk}(\varepsilon)$, which modifies the pristine band
structure $\varepsilon_{n\bk}$ of graphene described by the
\emph{noninteracting} Green's function, $G_{n\bk}^0(\varepsilon)=(\varepsilon -
\varepsilon_{n\bk} + i\eta)^{-1}$. While $G_{n\bk}^0$ is diagonal in the band
index, the disorder self-energy in~\eqref{eq:disavg} is, in general, not
diagonal and Eq.~\eqref{eq:disavg} must be solved by matrix inversion.

In the $T$-matrix approximation~~\cite{Rammer,Flensberg}, the impurity
self-energy is given by $\hat{\Sigma}_{\bk}(\varepsilon) = c_i
\hat{T}_{\bk\bk}(\varepsilon)$, where $c_i = N_i /N$ is the impurity
concentration (impurities per unit cell), and $\hat{T}_{\bk\bk}(\varepsilon)$
denotes the $\bk$-diagonal elements of the $T$ matrix. The $T$ matrix takes into
account multiple scattering off the individual impurities, and is given by the
integral equation
\begin{equation}
  \label{eq:Tmatrix} \hat{T}_{\bk\bk'}(\varepsilon) = \hat{V}_{\bk\bk'} +
\sum_{\bk''} \hat{V}_{\bk\bk''} \hat{G}_{\bk''}^0(\varepsilon)
\hat{T}_{\bk''\bk'}(\varepsilon) ,
\end{equation} where $\hat{V}_{\bk\bk'}$ are the impurity matrix elements.
The $T$-matrix self-energy is exact to leading order in the impurity
concentration $c_i$, and is therefore a good approximation for low impurity
concentrations, i.e., $c_i\ll 1$.

In the following, we obtain the GF and spectral function of Li@graphene based on
atomistic DFT (LCAO) calculations~\cite{calculations}~\nocite{GPAW,GPAW1,GPAW2}
of the band structure and impurity matrix elements sampled in the full BZ using
the atomistic method described in Refs.~\cite{Jauho:Symmetry,Kaasbjerg:First}
(see Ref.~\cite{Bernardi} for recent related developments).

\textbf{\textit{Li-adatom impurity potential.}}---For the initial
characterization of the Li adatoms, we have carried out standard DFT
calculations~\cite{calculations}, finding that the hollow site at the center of
the hexagon of the graphene lattice is the favored adsorption site at a distance
of $d=1.78\,\mathrm{\AA}$ above the graphene layer, and a net charge of $Q=-e
Z_\mathrm{Li}$ donated to the graphene lattice per Li adatom
($Z_\mathrm{Li}=+0.9$), consistent with previous
works~\cite{Cohen:First,Lichtenstein:Impurities}.

In our DFT calculated impurity matrix elements
$V_{\bk\bk'}^{nn'}=\bra{\psi_{n\bk}} \hat{V}_\mathrm{Li}
\ket{\psi_{n'\bk'}}$~\cite{calculations},
where $\ket{\psi_{n\bk}}$ is the Bloch state of the pristine system, the
microscopic details of the graphene-Li interaction are encoded in the DFT
Li-adatom impurity potential $\hat{V}_\mathrm{Li}$. For the sake of simplicity
we here express it as
\begin{equation}
    \label{eq:Vi} 
    \hat{V}_\mathrm{Li} = V_C(\hat{\br})  + \sum_{n\bk,n'\bk'} \ket{\psi_{n\bk}}
    V^{nn'}_{\mathrm{2s},\bk\bk'} \bra{\psi_{n'\bk'}}, 
\end{equation} 
where the two dominant contributions come from (1) the Coulomb potential
$V_C(\br)$ from the charged Li adatoms ($Q=+e Z_\mathrm{Li}$) which corresponds
approximately to a screened point-charge potential given by
$V_C(q,d) = e^2 Z_\text{Li} e^{-qd} /[2\epsilon_0 q \varepsilon(q)]$ in Fourier
space where $\varepsilon(q)$ is the two-dimensional (2D) static dielectric
function of graphene~\cite{Sarma:Carrier,screening}, and (2) the atomic Li 2s
state with energy $E_\mathrm{2s}$ which can be reduced to an effective potential
$V_{\mathrm{2s},\bk\bk'}^{nn'}$ described by the DFT
pseudopotential~\cite{Kaasbjerg:First} (see, e.g.,
Refs~\onlinecite{Lichtenstein:Adsorbates,Fabian:Resonant} for tight-binding
examples).
\begin{figure}[!t] 
  \centering
  \includegraphics[width=0.99\linewidth]{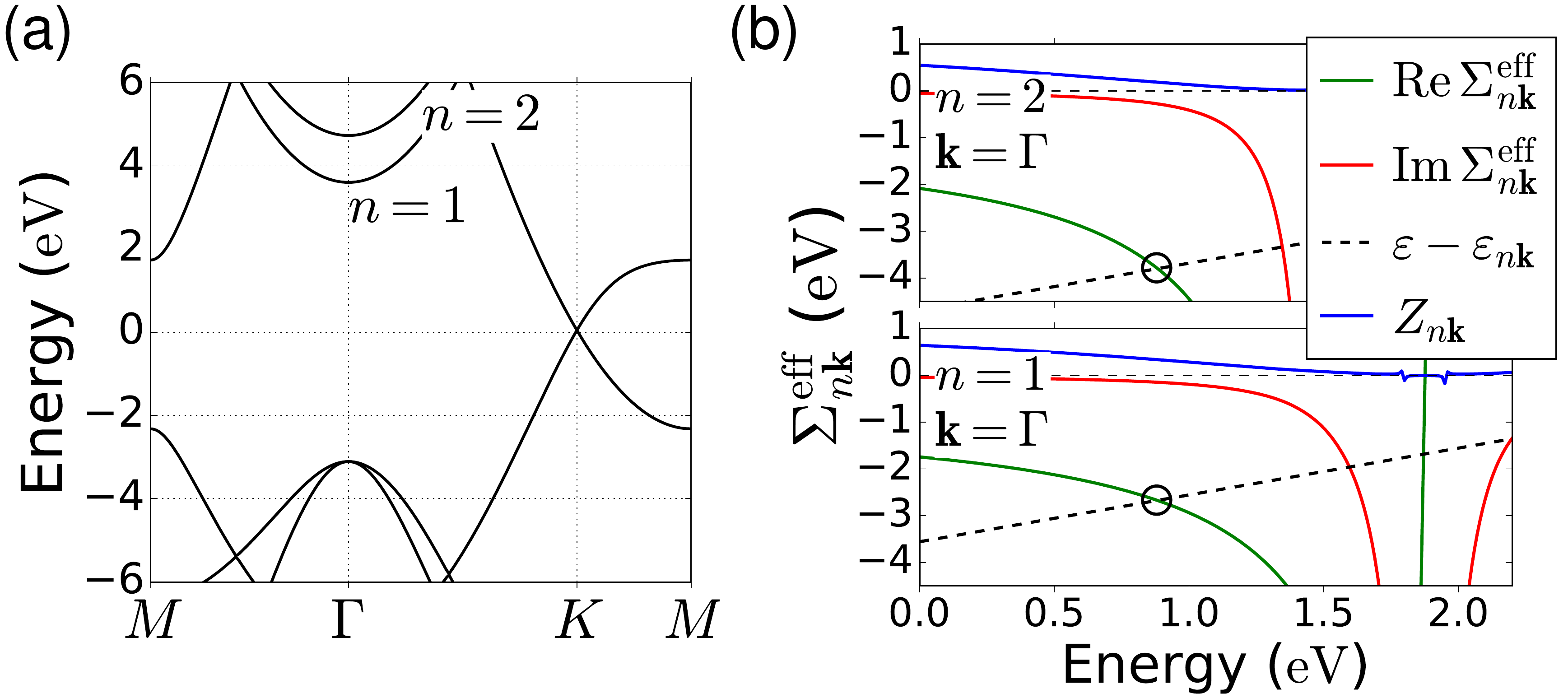}
  \caption{(a) Graphene bandstructure featuring two parabolic bands labeled
    $n=1,2$ which are associated with free-electron like surface
    states~\cite{Brandbyge:Simple}. (b) Diagonal elements of the real and
    imaginary parts of the effective self-energy in Eq.~\eqref{eq:Sigmaeff} for
    the $n=1,2$ surface states at $\bk=\bm{\Gamma}$ and
    $c_\mathrm{Li}=10\,\%$. The circles indicate the solutions to the QP
    equation corresponding to the hybrid impurity band in Li@graphene [cf. text
    below Eq.~\eqref{eq:spectral}].}
\label{fig:bandstructure}
\end{figure}

For an accurate description of the atomic 2s state in the $T$ matrix, we use a
Bloch-state basis $\{\ket{\psi_{n\bk}}\}$ which describes both the graphene
layer and the vacuum region where the Li atoms reside. In our DFT-LCAO based
method, this is achieved by introducing socalled "ghost" atoms in the surface
region which enlarge the standard LCAO basis for graphene. In the graphene band
structure shown in Fig.~\ref{fig:bandstructure}, this gives rise to the two
$\Gamma$-point centered parabolic bands (labeled $n=1,2$) located
$\sim 3.5$--5~eV above the Dirac point ($E=0$) which are absent in standard
tight-binding and DFT-LCAO calculations~\cite{Geim:RMP,Brandbyge:Simple}. The
two bands correspond to free-electron-like surface states located predominantly
outside the graphene plane with, respectively, even ($n=1$) and odd ($n=2$)
parity with respect to graphene's mirror symmetry plane~\cite{Brandbyge:Simple},
and they are instrumental for the occurrence of the impurity band in
Fig.~\ref{fig:intro}(b).

\textbf{\textit{Li@graphene spectral function.}}---In Fig.~\ref{fig:intro}(b) we
show the calculated spectral function
$A_\bk(\varepsilon)=\sum_n A_{n\bk}(\varepsilon)$ for $c_\mathrm{Li}=10\,\%$,
where $A_{n\bk}(\varepsilon) = -2 \mathrm{Im} \, G_{\bk}^{nn}(\varepsilon)$ is
given by the imaginary part of the diagonal elements of the GF. The dashed
horizontal line shows the Fermi level $E_F$ assuming a Li-induced carrier
density of $n=Z_\mathrm{Li} n_\mathrm{Li}$, where $n_\mathrm{Li}$ is the areal
density of Li atoms, and zero residual doping often present
experimentally~\cite{Damascelli:Evidence}. The spectral function exhibits two
distinct features which are absent in the pristine band structure of graphene.

The first is the appearance of a prominent $\Gamma$-centered parabolic impurity
band which starts $\sim 0.9$~eV above the Dirac point and extends up to
$\sim 2.5$~eV where it vanishes. As indicated by the Fermi level, this band is
populated at the considered $c_\mathrm{Li}$ and is hence important for the
electronic and transport properties of Li@graphene, including its potential
superconducting state~\cite{Damascelli:Evidence}. The second feature is a
pronounced renormalization and broadening of the conduction band near the $M$
point which is an indication of strong QP scattering.

As justified below, these features can be analyzed using a diagonal form of the
impurity-averaged GF,
$G_{n\bk}(\varepsilon)=[\varepsilon - \varepsilon_{n\bk} -
\Sigma_{n\bk}(\varepsilon)]^{-1}$.
Thus, the renormalized QP bands $\tilde{\varepsilon}_{n\bk}$ follow from the
solution to the QP equation
$\varepsilon - \varepsilon_{n\bk} - \mathrm{Re}\,\Sigma_{n\bk}(\varepsilon)=0$.
In the vicinity of $\tilde{\varepsilon}_{n\bk}$, the spectral function takes the
form
\begin{equation}
  \label{eq:spectral} 
  A_{n\bk}(\varepsilon) = Z_{n\bk} \frac{\gamma_{n\bk}}
  {(\varepsilon - \tilde{\varepsilon}_{n\bk})^2 + (\gamma_{n\bk}/2)^2} ,
\end{equation} 
where the wave-function renormalization, or QP weight, is given by
$Z_{n\bk}=[1- \partial_\varepsilon \mathrm{Re}\,
\Sigma_{n\bk}\vert_{\varepsilon=\tilde{\varepsilon}_{n\bk}}]^{-1}$,
and
$\gamma_{n\bk} = -2Z_{n\bk} \mathrm{Im} \,
\Sigma_{n\bk}\vert_{\varepsilon=\tilde{\varepsilon}_{n\bk}}$
is the linewidth broadening due to impurity scattering.

Besides renormalization of the pristine band structure, impurities with resonant
atomic levels may introduce new spectral features as is the case here. Such
features stem from additional solutions $\tilde{\varepsilon}_{\mathrm{imp},\bk}$
to the QP equation, and the spectral function acquires an additional impurity
component,
$A_{n\bk}(\varepsilon) \approx 2 \pi Z_{n\bk} \delta(\varepsilon -
\tilde{\varepsilon}_{n\bk}) + A_{n\bk}^\mathrm{imp}(\varepsilon)$,
where well-defined QPs are assumed and $A_{n\bk}^\mathrm{imp}$ is given by
Eq.~\eqref{eq:spectral} with
$\tilde{\varepsilon}_{n\bk} \rightarrow \tilde{\varepsilon}_{\mathrm{imp},\bk}$.
By virtue of the sum rule
$\int \tfrac{d\varepsilon}{2\pi} A_{n\bk}(\varepsilon)=1$, this results in a
reduction of the QP weights of the pristine bands from their unperturbed value
$Z_{n\bk}=1$.

In the rest of the paper, we clarify the microscopic origin of the
adatom-related features in Fig.~\ref{fig:intro}(b) as well as their dependence
on the concentration of Li adatoms.

\textbf{\textit{Hybrid impurity band.}}---The origin of the $\Gamma$-centered
impurity band can be traced back to the coupling between the $n=1,2$ surface
states in Fig.~\ref{fig:bandstructure} and the atomic Li 2s state via the last
term in Eq.~\eqref{eq:Vi}. This introduces a pole in the $T$ matrix at the
renormalized energy $\tilde{E}_{2\mathrm{s}} \approx 2.55$~eV of the 2s
state. At $\bk\sim \bm{\Gamma}$, the pole together with the broken mirror plane
symmetry in Li@graphene gives rise to large diagonal and off-diagonal $T$-matrix
elements between the $n=1,2$ surface states~\cite{supplemental}, which dominate
all other elements. To facilitate a simple analysis, we can thus approximate the
GF in the $n=1,2$ surface-state subspace by the inverse of its $2\times 2$
subblock on the right-hand side of Eq.~\eqref{eq:disavg}. The diagonal elements
of the GF take the form
$G_{n\bk}(\varepsilon)=[\varepsilon - \varepsilon_{n\bk} -
\Sigma_{n\bk}^\mathrm{eff}(\varepsilon)]^{-1}$,
where the \emph{effective} self-energy,
\begin{equation}
    \label{eq:Sigmaeff} \Sigma_{n\bk}^{\text{eff}}(\varepsilon) =
\Sigma_\bk^{nn}(\varepsilon) +
\frac{\Sigma_\bk^{n\bar{n}}(\varepsilon)\Sigma_\bk^{\bar{n}n}(\varepsilon)}
{\varepsilon - \varepsilon_{\bar{n}\bk} -
\Sigma_\bk^{\bar{n}\bar{n}}(\varepsilon)} , \quad \bar{n}\neq n ,
\end{equation} 
describes virtual transitions between the surface state $n$ and the Li 2s state
either (i) directly (first term), or (ii) via the other surface state $\bar{n}$
(second term).
\begin{figure}[!t] 
  \centering
  \includegraphics[width=0.99\linewidth]{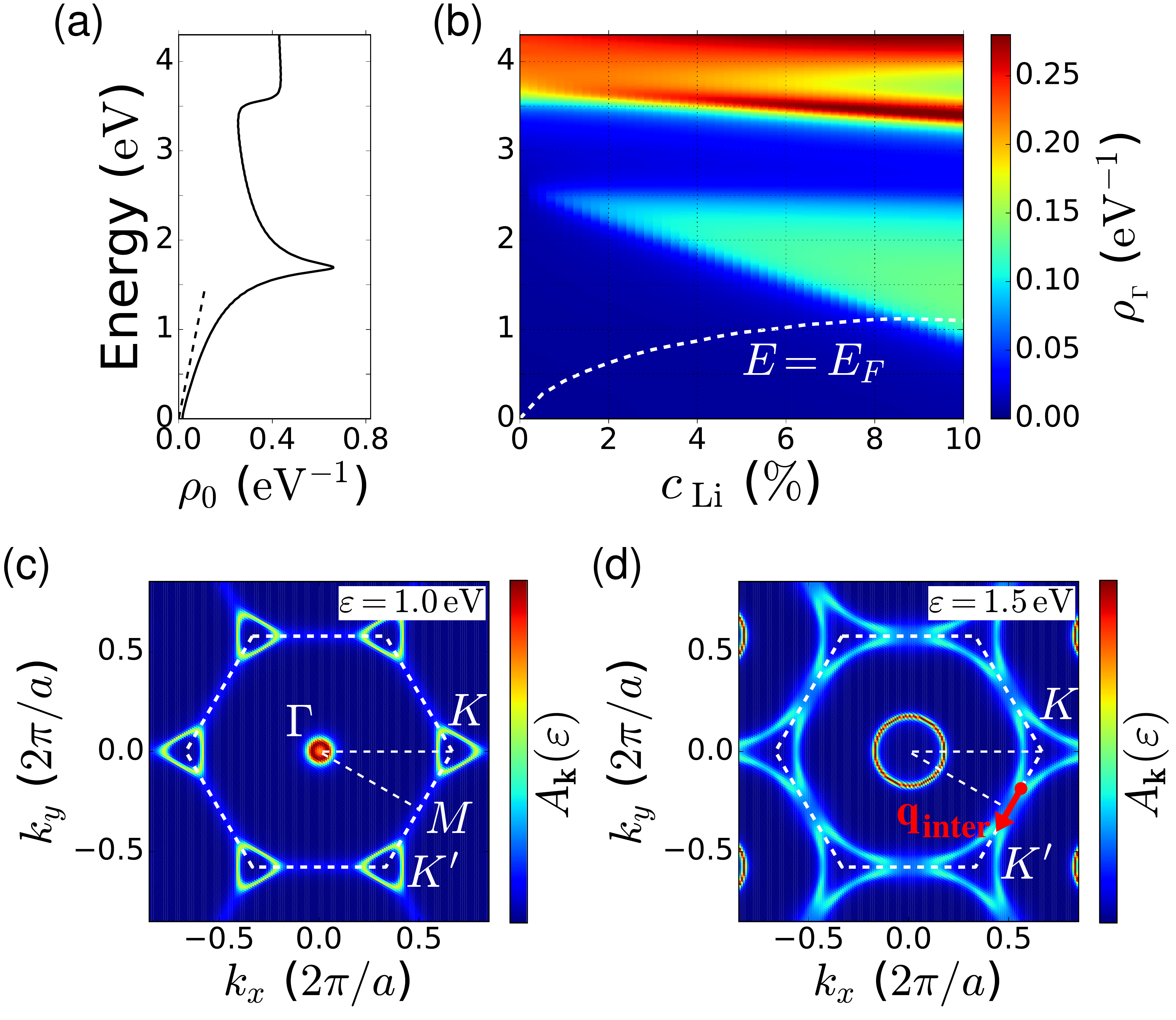}
  \caption{(a) Density of states of pristine graphene with the dashed line
    showing the linear DOS in the Dirac model. (b) DOS of the $\Gamma$-centered
    bands in Li@graphene as a function of the Li concentration and energy. The
    dashed line denotes the Li-induced Fermi level assuming zero residual
    doping. The constant DOS which develops with increasing $c_\mathrm{Li}$ in
    the energy range $\sim 0.9$--$2.5$~eV is due to the hybrid impurity band in
    Fig.~\ref{fig:intro}(b). (c) and~(d) Spectral function at, respectively,
    $\varepsilon=1.0$~eV and $\varepsilon=1.5$~eV for $c_\mathrm{Li}=10\,\%$. In
    (d), the strong trigonal warping results in a large reduction of the
    intervalley scattering wave vector (red arrow).}
\label{fig:spectral}
\end{figure}

With this form of the diagonal elements of the GF, the analysis in
Eq.~\eqref{eq:spectral} applies. In Fig.~\ref{fig:bandstructure}(b) we show the
real (green lines) and imaginary parts (red lines) of the diagonal elements of
the effective self-energy, $\Sigma_{n\bk}^{\text{eff}}$, together with the
calculated QP weight $Z_{n\bk}$ (blue lines) and
$\varepsilon - \varepsilon_{n\bk}$ (black dashed). The intersection of the
latter with $\mathrm{Re}\Sigma_{n\bk}^\mathrm{eff}$ (marked with circles)
signifies the emergence of the $\Gamma$-centered hybrid impurity band in
Fig.~\ref{fig:intro}(b). The QP weights at the $\Gamma$ point are, respectively,
$Z_{1\Gamma}\approx 0.3$ and $Z_{2\Gamma}\approx 0.2$, yielding a total QP
weight of $Z_{\mathrm{imp},\Gamma} \sim 0.5$ for the impurity band.

We have thus identified the spectral $\Gamma$-point feature observed in
ARPES~\cite{Damascelli:Evidence} as a low-energy hybridized impurity band
arising from the coherent coupling between the surface states and the Li 2s
state as described by Eq.~\eqref{eq:Sigmaeff}. In the periodic LiC$_6$
structure, the analog of this band plays a pivotal role for the predicted
superconductivity by enhancing the el-ph coupling at the Fermi
level~\cite{Mauri:Phonon}.

It is therefore interesting to investigate the concentration dependence of the
impurity band and its alignment with the Fermi level in Li@graphene. In
Figs.~\ref{fig:spectral}(a) and~\ref{fig:spectral}(b), we show the DOS (per unit
cell)
$\rho(\varepsilon) = -\tfrac{1}{N\pi} \mathrm{Im}\,[\mathrm{Tr}\,
\hat{G}_\bk(\varepsilon)]$
for, respectively, (i) pristine graphene ($\rho_0$), and (ii) the
$\Gamma$-centered bands in Li@graphene ($\rho_\Gamma$) obtained by restricting
the $\bk$ sum in the trace to a region around the $\Gamma$ point enclosing the
relevant bands. The dashed line in Fig.~\ref{fig:spectral}(b) marks the position
of the Fermi level corresponding to the Li-induced carrier density
$n=Z_\mathrm{Li}n_\mathrm{Li}$. In the Dirac model the Fermi energy scales as
$E_F = \hbar v_F \sqrt{\pi Z_\mathrm{Li}n_\mathrm{Li}} \approx 120 \,
\sqrt{n/(10^{12}\,\mathrm{cm}^{-2})}\;\mathrm{meV}$,
whereas the Fermi energy in Fig.~\ref{fig:spectral}(b) deviates from this
square-root dependence at high $c_\mathrm{Li}$ where it flattens out due to the
population of the hybrid band and the nonlinear part of the Dirac cone. The
figure illustrates the development of the hybrid impurity band which starts from
the position of the Li 2s state $\tilde{E}_{2s}$ and moves down towards the
Dirac cone with increasing Li concentration. Interestingly, our calculations
show that the hybrid impurity band aligns with the Fermi level on the Dirac cone
at Li concentrations as low as $c_\mathrm{Li}\sim 8\,\%$ ($E_F \approx 1.1$~eV)
where the DOS $\rho_\Gamma$ of the impurity band and the Dirac-cone DOS $\rho_0$
are comparable. In the presence of a residual doping of
graphene~\cite{Damascelli:Evidence}, this situation is realized at even lower
$c_\mathrm{Li}$.

\textbf{\textit{Dirac-cone QP properties.}}---Finally, we consider the effect of
adatom-induced QP scattering on the renormalization and linewidth broadening of
the Dirac-cone bands in Fig.~\ref{fig:intro}(b) and Figs.~\ref{fig:spectral}(c)
and~\ref{fig:spectral}(d). As the self-energy on the Dirac cone is
diagonal~\cite{supplemental}, the renormalized QP bands
$\tilde{\varepsilon}_{n\bk}$ and linewidth broadening $\gamma_{n\bk}$ can be
obtained as explained above and below Eq.~\eqref{eq:spectral}.

\begin{figure}[!t] \centering
  \includegraphics[width=0.99\linewidth]{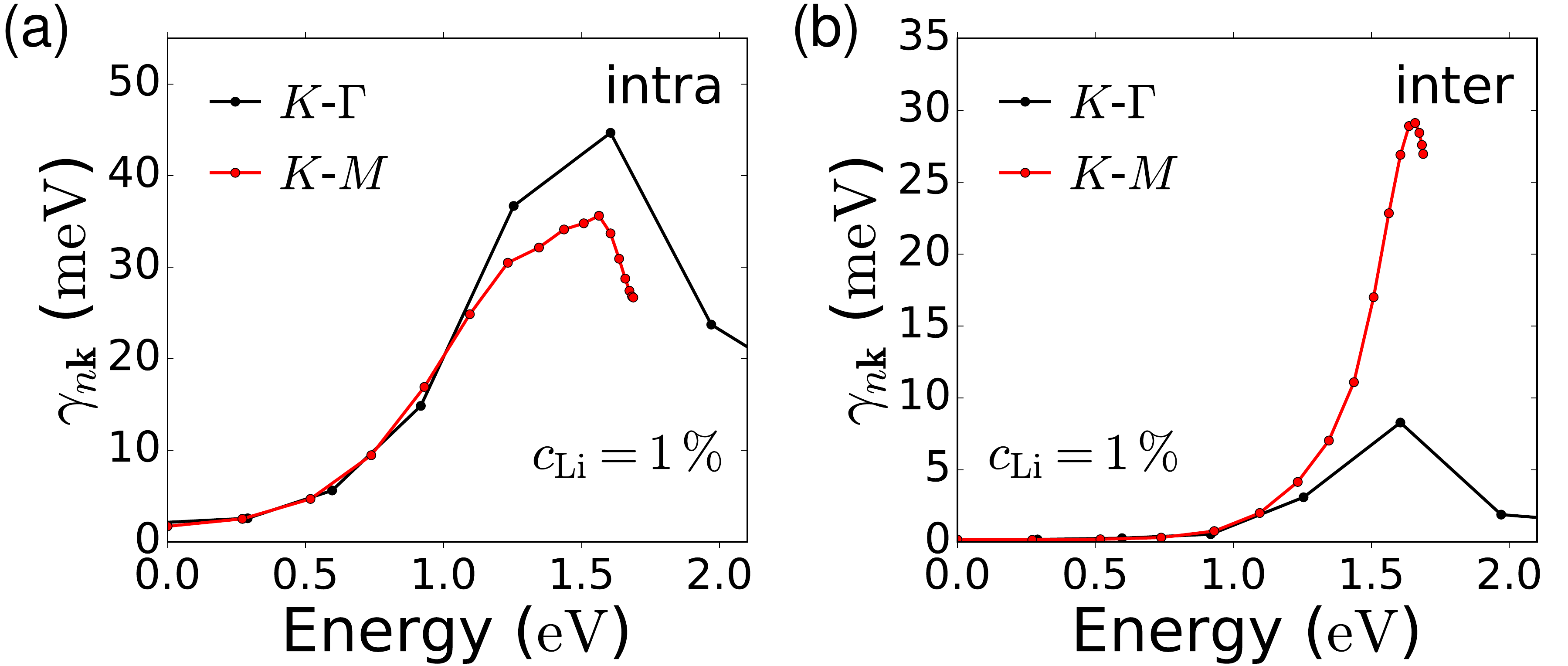}
  \caption{Adatom-induced linewidth broadening $\gamma_{n\bk}$, or QP scattering
    rate $\tau_{n\bk}^{-1}=\gamma_{n\bk}/\hbar$, in the conduction band as a
    function of the on-shell energy $\tilde{\varepsilon}_{n\bk}$ (along the
    indicated BZ paths) at $c_\mathrm{Li}=1\,\%$. The total broadening has been
    split up into contributions from (a) intravalley and (b) intervalley
    scattering, respectively. The peaks in the linewidth broadening correlate
    with the position of the vHS in the DOS in Fig.~\ref{fig:spectral}(a).}
\label{fig:scattering}
\end{figure}

In Fig.~\ref{fig:scattering} we show the energy dependence of the linewidth
broadening of the conduction band along the $K$-$\Gamma$ and $K$-$M$ paths for a
fixed Li concentration of $c_\mathrm{Li}=1\,\%$ at which the impurity band is
absent [cf. Fig.~\ref{fig:spectral}(a)]. For clarity, we have separated out the
contributions from intravalley (left) and intervalley (right) scattering using
the optical theorem~\cite{supplemental,Kaasbjerg:First}. At energies below
$\sim 1\,$eV, the intravalley rate exceeds the intervalley rate by far. This is
consistent with charged-impurity scattering where intervalley scattering with
$q\approx \abs{\bK -\bK'}$ is weak due to the $q$ dependence of the 2D Coulomb
potential $V_C(q,d)$. We find no indications of resonant
scattering~\cite{Basko:Resonant,Falko:Adsorbate,Katsnelson:Resonant,Fabian:Resonant}
which is suppressed by the remote energy $\tilde{E}_\mathrm{2s}$ of the Li 2s
state as well as the invisibility of short-range impurity potentials due to
adatoms in the hollow
site~\cite{Ferreira:Impurity,Silva:Symmetry,Fabian:Resonant}. At higher
energies, both the intra and intervalley rates increase dramatically and peak at
the energy $\varepsilon\approx 1.75$~eV of the van Hove singularity (vHS) in the
DOS in Fig.~\ref{fig:spectral}(a). Whereas the increasing DOS at the vHS is the
main reason for the increasing intravalley rate, also the strong trigonal
warping of the Dirac cones seen in Figs.~\ref{fig:spectral}(c)
and~\ref{fig:spectral}(d) is important in order to explain the increase in the
intervalley rate. As illustrated in Fig.~\ref{fig:spectral}(d), the trigonal
warping reduces the intervalley scattering wave vector markedly and thereby
enhances the scattering probability due to the $q$ dependence of Coulomb
potential. This effect is most pronounced on the $K$-$M$ path, where the
reduction in the $K\rightarrow K'$ intervalley wave vector is strongest, and
results in an anisotropic linewidth broadening also visible on the
$\varepsilon = 1.5$~eV constant-energy contours in
Fig.~\ref{fig:spectral}(d). In Figs.~\ref{fig:spectral}(c) and
\ref{fig:spectral}(d), $c_\mathrm{Li}=10\,\%$, and $\Gamma \leftrightarrow K$
scattering between the overlapping impurity band and Dirac cones may also
contribute to the broadening. However, the narrow linewidth of the impurity band
shows that this scattering channel is weak because of a small impurity matrix
element between the spatially separated surface and Dirac-cone states.
Dirac-cone QP scattering is thus not affected markedly by the impurity band.

Interestingly, the strong peak in the linewidth broadening at the vHS is
accompanied by a pronounced renormalization of the conduction band in the same
energy range as demonstrated by the calculated QP band in
Fig.~\ref{fig:intro}(b) (solid blue line). Along the $K$-$M$ path, the
conduction band exhibits a pronounced downshift and a kink $\sim 100$--$200$~meV
below the vHS, resembling experimental ARPES features in highly
adatom-doped~\cite{Rotenberg:Extended,Gruneis:Observation} as well as
intercalated~\cite{Starke:Introducing} graphene. At high doping
levels~\cite{Rotenberg:Extended,Gruneis:Observation,Hofmann:Elph}, this
adatom-induced kink may interfere with el-ph related kinks located at the
optical phonon energy $\sim 200$~meV below the Fermi
level~\cite{Mauri:ElPh,Louie:Velocity,Sarma:Phonon,Louie:Angle}, thus obscuring
the analysis of the el-ph interaction~\cite{Louie:Van,Mauri:Electronic} in the
regime relevant for superconductivity.

\textbf{\textit{Conclusions}.}---We have studied the spectral function and
quasiparticle scattering in disordered Li-decorated graphene with an atomistic
$T$-matrix method. We demonstrated that (i) the experimentally observed
low-energy spectral feature at the $\Gamma$ point~\cite{Damascelli:Evidence}
originates from a Li-dependent hybrid impurity band which aligns with the
Li-induced Fermi level at $c_\mathrm{Li}\approx 8\,\%$, and (ii) Li-induced
charged-impurity scattering produces a strong linewidth broadening and a
concomitant downshift and kink in the conduction band in the vicinity of the van
Hove singularity in the graphene DOS. Our findings are highly relevant for
future studies of
transport~\cite{Ishigami:Charge,Fuhrer:Correlated,Forti:Alkali,Eisenstein:Transport}
and ARPES~\cite{Rotenberg:Extended,Gruneis:Observation,Starke:Introducing} as
well as analyses of the el-ph interaction~\cite{Louie:Van,Mauri:Electronic} and
superconductivity~\cite{Mauri:Phonon,Damascelli:Evidence} in adatom-doped
graphene.

\begin{acknowledgments} \textbf{\textit{Acknowledgments}.}---We would like to
  thank T. Olsen, M. Brandbyge and J.~A. Folk for fruitful discussions and
  comments. K.K. acknowledges support from the European Union's Horizon 2020
  research and innovation programme under the Marie Sklodowska-Curie Grant
  Agreement No.~713683 (COFUNDfellowsDTU). The Center for Nanostructured
  Graphene (CNG) is sponsored by the Danish National Research Foundation,
  Project DNRF103.
\end{acknowledgments}

\bibliographystyle{apsrev}
\bibliography{main}

\begin{thebibliography}{56}
\bibitem [{cal()}]{calculations}%
  \BibitemOpen
  \href@noop {} {}\bibinfo {note} {All DFT calculations have been performed
  with the electronic structure code GPAW~\cite{GPAW,GPAW1,GPAW2} within the
  projector augmented-wave method, using the LDA xc functional and a LCAO
  double-zeta polarized (DZP) basis set. The ground-state calculation of the
  optimized structure and Bloch functions was performed using a $21\times 21$
  $\bk$-point sampling of the BZ. The Li impurity potential was obtained using
  a $11 \times 11$ supercell and a $3\times 3$ $\bk$-point sampling. All
  calculations include 10~{\AA} of vacuum between graphene and the cell
  boundaries in the vertical direction imposed with Dirichlet boundary
  conditions. The impurity matrix elements, $T$-matrix self-energy and GF were
  obtained on a $54\times 54$ BZ grid, including the 12 lowest bands and using
  a numerical broadening of $\eta=100$~meV.}\BibitemShut {Stop}%
\bibitem [{scr()}]{screening}%
  \BibitemOpen
  \href@noop {} {}\bibinfo {note} {Here, screening of the impurity potential by
  the Li-induced carriers is accounted for in the supercell DFT calculation of
  the impurity potential corresponding to a Fermi energy of
  $E_F=0.5$~eV.}\BibitemShut {Stop}%
\end{thebibliography}

\pagebreak
\widetext
\clearpage


\section*{Supplementary material}

# Supplemental material for "Signatures of adatom effects in the quasiparticle spectrum of Li-doped graphene"

Kristen Kaasbjerg[*] and Antti-Pekka Jauho
*Center for Nanostructured Graphene (CNG), Department of Physics,
Technical University of Denmark, DK-2800 Kgs. Lyngby, Denmark*
(Dated: November 19, 2019)

## S1. $T$-MATRIX SELF-ENERGY

Here, we include plots in Fig. S1 of the calculated $T$-matrix self-energy,

$$\hat{\Sigma}_{\mathbf{k}}(\varepsilon) = c_i \hat{T}_{\mathbf{kk}}(\varepsilon), \tag{S1}$$

to show that it indeed has the properties mentioned in the main text.

### A. Surface-state subspace: Li 2s state

The left panel in Fig. S1 shows the $T$-matrix self-energy in the subspace of the two surface states corresponding to the $n=1,2$ bands in Fig. 2(a) of the main text. In this subspace, the self-energy is a dense matrix with almost identical diagonal and off-diagonal elements, and the signature of a pole at $\sim 2.55$ eV due to the coupling to the Li 2s state clearly emerges.

### B. Dirac-cone subspace: Li-induced impurity scattering

The right panel in Fig. S1 shows the $T$-matrix self-energy in the subspace of the valence ($v$) and conduction ($c$) bands on the Dirac cone. Here, the self-energy is dominated by impurity scattering off the Li-induced Coulomb potential which results in a diagonal matrix structure and a pronounced increase in the imaginary part at the van Hove singularities in the DOS.

The wiggles on curves visible at low and high energies are due to the discrete **k**-point sampling of the Brillouin zone and demonstrate the necessity of using a numerical broadening $\eta$ which matches the energy spacing between neighboring **k** points in the band structure.

In both of the above cases, there are small matrix elements to other remote bands. However, due to the large separation in energy they can safely be neglected in our simplified analysis of the spectral features. The excellent

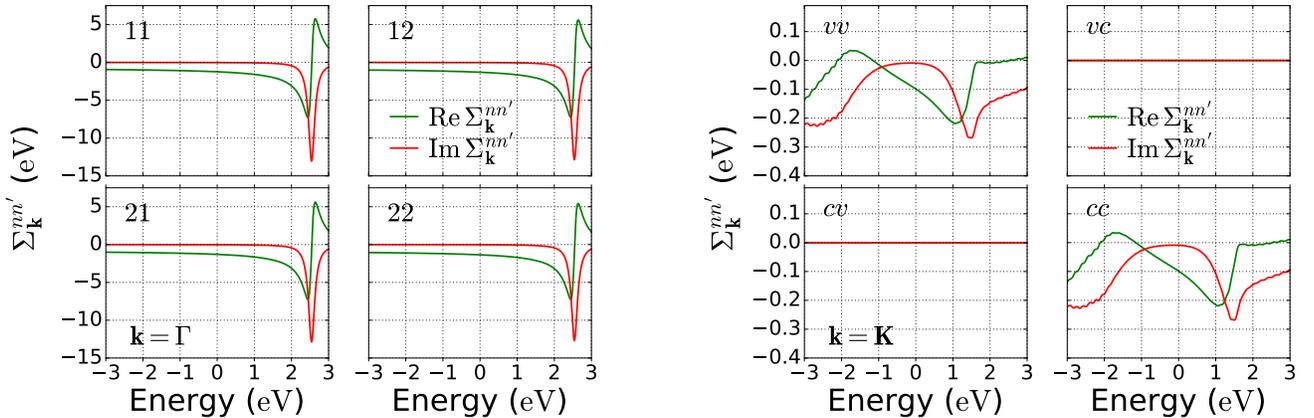

FIG. S1. Matrix structure of the $T$-matrix self-energy for Li@graphene. (Left) Self-energy in the $2 \times 2$ subspace spanned by the surface-state bands ($n=1,2$). (Right) Self-energy between the valence ($v$) and conduction ($c$) bands of the Dirac cone ($n=v,c$). Parameters: $c_{\text{Li}} = 10\%$, $\eta = 100$ meV.

agreement between the spectral function in Fig. 1(b) of the main manuscript obtained from the full matrix GF and our simplified analysis [Fig. 1(b) solid blue lines + Fig. 2(b)] confirms this.

## S2. OPTICAL THEOREM

The individual contributions to the linewidth broadening from intra- and intervalley scattering can be obtained by taking advantage of the optical theorem by which the diagonal elements of the imaginary part of the $T$ matrix can be expressed as

$$-2\mathrm{Im}T_{\mathbf{kk}}^{nn}(\varepsilon) = -2\mathrm{Im}\sum_{n'\mathbf{k}'}\frac{|T_{\mathbf{kk}'}^{nn'}(\varepsilon)|^2}{\varepsilon - \varepsilon_{n'\mathbf{k}'} + i\eta}$$
$$= 2\pi\sum_{n'\mathbf{k}'}|T_{\mathbf{kk}'}^{nn'}(\varepsilon)|^2\delta(\varepsilon - \varepsilon_{n'\mathbf{k}'}). \tag{S2}$$

Here, the sums run over possible transitions $n\mathbf{k} \to n'\mathbf{k}'$ with the scattering amplitude given by the off-diagonal elements $T_{\mathbf{kk}'}^{nn'}$ of the $T$ matrix. By splitting the $\mathbf{k}'$ sum into sums over intra- and intervalley processes, $\sum_{\mathbf{k}'} \to \sum_{\mathbf{k}'\in\mathrm{intra}} + \sum_{\mathbf{k}'\in\mathrm{inter}}$, the contributions to the lifetime broadening from intra- and intervalley scattering can be extracted.



\end{document}